\documentclass[fleqn,usenatbib]{mnras}

\usepackage{newtxtext,newtxmath}

\usepackage[T1]{fontenc}

\DeclareRobustCommand{\VAN}[3]{#2}
\let\VANthebibliography\thebibliography
\def\thebibliography{\DeclareRobustCommand{\VAN}[3]{##3}\VANthebibliography}

\usepackage{graphicx}	
\usepackage{amsmath}	

\title[RPS radio tails of Shapley supercluster galaxies]{Ram-pressure
  stripped radio tails detected in the dynamically active
  environment of the Shapley Supercluster}

\author[Merluzzi et al.]{P. Merluzzi$^{1}$\thanks{E-mail: paola.merluzzi@inaf.it}, T. Venturi$^{2,3}$,
  G. Busarello$^{1}$, G. Di Gennaro$^{4,2}$, S. Giacintucci$^{5}$, V. Casasola$^2$,
 \newauthor D. Krajnovi\'c$^{6}$, T. Vernstrom$^{7}$, E. Carretti$^{2}$,
 O. Smirnov$^{3,8,2}$, K. Trehaeven$^{7,3,2}$, C.~S. Anderson$^{9}$,
 \newauthor J. Chesters$^{10}$, G. Heald$^{10}$, A.~M. Hopkins$^{11}$, B. Koribalski$^{12,13}$
\\
$^1$ INAF-Osservatorio Astronomico di Capodimonte, salita Moiariello 16, I-80131 Napoli, Italy \\
$^2$ INAF-Istituto di Radioastronomia, via Gobetti 101, I-40129 Bologna, Italy \\
$^3$ Center for Radio Astronomy Techniques and Technologies, Rhodes University,Grahamstown 6140, South Africa\\
$^4$ Hamburger Sternwarte, Universit\"at Hamburg, Gojenbergsweg 112, 21029 Hamburg, Germany\\
$^5$ Naval research Laboratory, 4555 Overlook Avenue SW, Code 7213, Washington, DC 20375, USA\\
$^6$ Leibniz-Institut f\"ur Astrophysik Potsdam (AIP), An der Sternwarte 16, 14482, Potsdam, Germany\\
$^7$ ICRAR, The University of Western Australia, 35 Stirling Hwy, 6009, Crawley, Australia \\
$^8$ South African Radio Astronomy Observatory, 2 Fir Street, Black River Park, Observatory, Cape Town 7925, South Africa\\
$^{9}$ Research School of Astronomy \& Astrophysics, The Australian National University, Canberra ACT 2611, Australia \\
$^{10}$ CSIRO, Space and Astronomy, PO Box 1130, Bentley, WA 6102, Australia \\
$^{11}$ School of Mathematical and Physical Sciences, 12 Wally's Walk Macquarie University, NSW 2109, Australia \\
$^{12}$ Australia telescope National Facility, CSIRO Astronomy and Space Science, P.O. Box 76, Epping, NSW 1710, Australia \\
$^{13}$ School of Science, Western Sydney University, Locked Bag 1797, Penrith, NSW 2751, Australia 
}

\date{Accepted XXX. Received YYY; in original form ZZZ}

\pubyear{2015}

\begin{document}
\label{firstpage}
\pagerange{\pageref{firstpage}--\pageref{lastpage}}
\maketitle

\begin{abstract}
We study the radio continuum emission of four galaxies experiencing
ram-pressure stripping in four clusters of the Shapley supercluster at
redshift $z\sim 0.05$. Multi-band (235-1367\,MHz) radio data,
complemented by integral-field spectroscopy, allow us to detect and
analyse in detail the non-thermal component both in the galaxy discs
and the radio continuum tails. Three galaxies present radio continuum
tails which are tens of kiloparsecs long.  By deriving the radio
spectral index in the inner and outer tails and comparing our findings
with the distribution of the extraplanar ionised gas and the results
of $N$-body/hydrodynamical simulations, we demonstrate that these
tails are caused by the ram pressure which, together with the ionised
gas, sweeps the magnetic field from the galaxy discs. We suggest that
the radio continuum emission in these tails can be differently powered
by ({\it i}) {\it in situ} star formation; ({\it ii}) relativistic
electrons stripped from the disc; ({\it iii}) shock excitation or a
combination of them. All the ram-pressure stripped galaxies are found
in environments where cluster-cluster interactions occurred and/or are
ongoing thus strongly supporting the thesis that cluster and group
collisions and mergers may locally increase the ram pressure and
trigger hydrodynamical interactions between the intracluster medium
and the interstellar medium of galaxies.

\end{abstract}




\section{Introduction}

Galaxies orbiting in the cluster environment experience the
hydrodynamical interaction of their cold interstellar medium (ISM)
with the hot diffuse intracluster medium (ICM). This interplay may
result in the ISM evaporation \citep{CS77} and `starvation'
\citep{LTC80} or may remove the cold gas supply by means of the
external pressure exerted by the ICM on the ISM \citep[][]{GG72}.  The
primary strength of these processes consists of explaining the lower
star formation rates \citep[SFRs; e.g.][]{BNM00} and the redder
colours \citep[e.g.][]{BNB09} as well as the H{\sc I} deficiency
\citep[e.g.][]{GR-V85,BVC01} seen in cluster galaxies with respect to
the field population. At present, galaxy evolution studies support the
ram-pressure stripping (RPS) as the dominant hydrodynamical mechanism
contributing to the evolution of cluster galaxies \citep[for a review
  see][and references therein]{BFS22}.

In their pioneering work \citet{GG72} modelled the RPS process by
means of an analytic expression where the ram pressure opposes the
gravitational restoring force per unit area - a decreasing function of
the disc radius. It results that in the dense environment of the
cluster cores the ram pressure can effectively remove the cooler ISM
in the galaxy starting from outside and thus quenching star formation
in the ram-pressure stripped regions. Subsequent studies based on
observations \citep[e.g.][]{ShaSSIII,Brown+17,Vulcani+21,Roberts+21II}
and simulations \citep[e.g.][]{MBD03,RH05,Bekki09} showed that the RPS
can occur, or start at least, also in less dense environments and/or
out of the cluster cores differently affecting high- and low-mass
galaxies. The time-scales for RPS is about one cluster crossing time
($\sim10^9$\,yr).

The characteristic signatures of RPS are the presence of {\it i})
extraplanar ISM stripped from the galaxy disc in the shape of one-side
tail; {\it ii}) distortion and ultimate truncation of the gaseous disc
without corresponding distortion of the old stellar component
\citep{BG06}.  Besides, the ISM of late-type galaxies has a
multi-phase nature whose components (warm diffuse ionised hydrogen,
cold and warm neutral hydrogen, dense molecular gas, interstellar
dust), having different surface densities, are differently affected by
the ram pressure
\citep[see][]{Casasola+04,Corbelli+12,Boselli+14,Boselli+18,Davies+19,Poggianti+19,Brown+21}. Moreover,
as observed in nearby clusters
\citep[e.g.][]{Gavazzi+95,Miller+09,Murphy+09,Vollmer+04,Vollmer+10,Vollmer+13},
also the non-thermal component can be stripped from the galaxy discs
forming cometary tails. Actually, the first case of late-type galaxies
experiencing RPS was detected in the radio continuum
\citep{Gavazzi78}.

It follows that multi-band observations of both the tail and the
galaxy are crucial to distinguish among the different components in
the tail, and high-resolution observations are required in order to
resolve the physics of the gas, the stars and the plasma in the
galaxy. Moreover, multi-band studies of individual galaxies affected
by RPS are often complemented with dedicated numerical simulations
\citep[see][]{Vollmer+01,Vollmer+18,ACCESSV,ShaSSIII,Bellhouse+21}

The effectiveness of such cluster-related mechanism in transforming
galaxies encouraged an extensive investigation of RPS supported by the
capabilities of the new observing facilities and the availability of
focused and highly resolved simulations down to scales of few tens of
parsecs. The spectacular appearance of an ongoing RPS also plays a
role in the general interest to study the RPS process and contributed
to boost systematic searches of RPS candidates, mainly based on
visible photometry \citep{ESE14,McP+16,P+16}, but recently also using
radio continuum \citep[e.g.][see
  below]{Roberts+21I,Lal+22,Roberts+23}. These works enabled to extend
the census of ongoing RPS in nearby clusters and to carry out detailed
studies of individual galaxies exploiting the now available
instrumental capabilities of the integral-field spectrographs
\citep[e.g.][]{ACCESSV,ShaSSIII,Fumagalli+14,Fossati+16,Fossati+19,Consolandi+17,GASPI-17,GASPXV-19}.

The advent of the new generation radio telescopes has given new
impetus to the study of galaxy tails in radio continuum allowing their
detection at sensitivities and resolutions which were inaccessible to
wide-field radio surveys until now. Through 120-168\,MHz radio
continuum images from the LOFAR two-meter Sky Survey,
\citet{Roberts+21I,Roberts+21II} identified star-forming galaxies
($\sim 150$) with extended asymmetric radio continuum tails likely
experiencing RPS in low redshift ($z< 0.05$) clusters and
groups. \citet{Chen+20} carried out a 1.4\,GHz continuum and H{\sc I}
emission survey in the Coma cluster to investigate the radio
properties of RPS galaxies. They detected radio continuum tails in
50\% of the targets demonstrating the widespread presence of
relativistic electrons and magnetic fields in the stripped
tails. \citet{Lal+22} extended the study of the Coma cluster using
uGMRT detecting 19 RPS radio continuum tails out of the observed 24
extended radio sources which are cluster members. With the collected
uGMRT observations, they were able to derive the spectral structure in
all the tailed radio sources. Thanks to the new instrumentation
facilities, we can now address the open issues about the nature and
origins of the radio tail in RPS galaxies and how the radio continuum
emission in the tails correlates with the emission in other bands and
with the galaxy environment. Sophisticated simulations including the
effects of either intra-cluster or galaxy magnetic fields are also
available \citep[e.g.][]{Rus+14,TS14}.

\begin{figure}
\begin{centering}
\includegraphics[width=85mm]{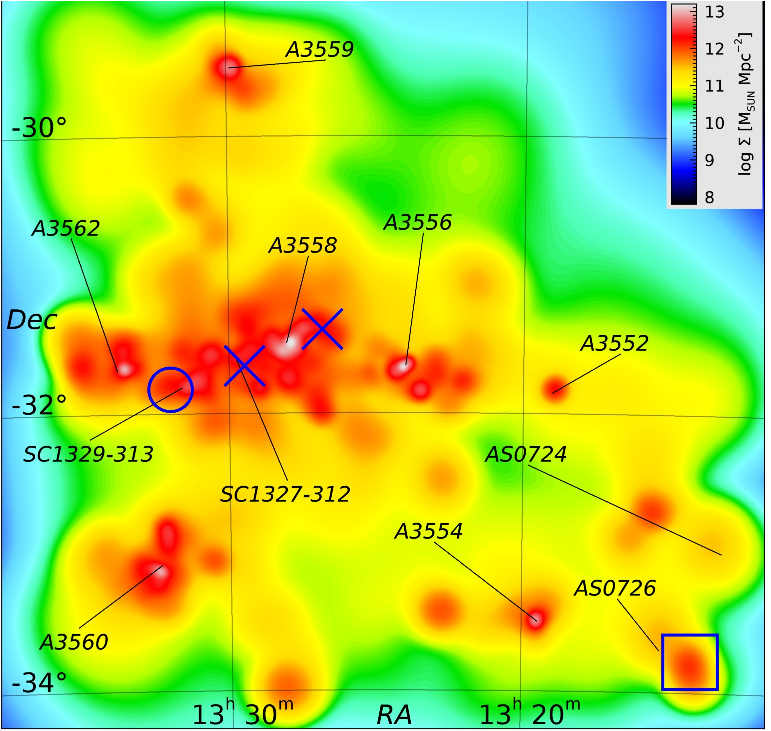}  
\end{centering}
\caption{ The Shapley Supercluster Survey stellar mass surface density
  (M$_\odot$\,Mpc$^{-2}$) derived from the WISE flux at 3.4\,$\mu$m
  \citep{ShaSSI}. Abell clusters and groups are labeled by black
  straight lines pointing on the X-ray centre for all
  systems. Different symbols indicate the positions of the RPS
  galaxies. SOS\,61086 \citep[circle, see][]{Venturi+22}, SOS\,90630
  and SOS\,114732 (crosses, their radio emission is analysed in this
  work) are covered by ASKAP, MeerKAT, GRMT and uGMRT imaging, while
  for ShaSS\,421 (box, presented in this work), ASKAP imaging only is
  currently available (see text).
\label{fig:gals}}
\end{figure}

In this framework, the role of the cluster and/or group interactions
in triggering the RPS mechanism is debated
\citep[see][]{Owers+12,ShaSSIII,EK19,Venturi+22}. Deep and
high-resolution radio continuum imaging of merging clusters provides,
on one side, detailed descriptions of the cluster merger features
tracing the intra-cluster diffuse non-thermal emission, on the other
side, it reveals the radio continuum tail of individual galaxies. This
allows us to investigate the cause-effect relationship, i.e. the link
between the large-scale mass assembly and the involved galaxies.

Superclusters are the most massive structures in the Universe
($10^{15}-10^{16}$\,M$_\odot$). In such dynamically active and locally
dense structures one can potentially observe the effects of dynamical
processes such as cluster-cluster collisions and group-cluster mergers
and sample different environments from cluster cores to filaments and
fields. In the harsh environment of the Shapley supercluster, we have
undertaken a search of supercluster members showing radio continuum
tails with the aim to investigate if these features are linked with
the ongoing or past cluster-cluster interaction.

\begin{table*}
  \centering
\caption{Properties of RPS galaxies}
{\small
\begin{tabular}{lcccc}
\hline
{\bf Property}  & {\bf SOS\,61086} & {\bf SOS\,90630} & {\bf SOS\,114372} & {\bf ShaSS\,421} \\
\hline
{\bf Coordinates} &&&& \\
ShaSS$^1$ & 13\,31\,59.80\,-31\,49\,22.2 & 13\,29\,28.53\,-31\,39\,25.6 & 13\,26\,49.79\,-31\,23\,44.3 & 13\,14\,07.43\,-33\,46\,21.1 \\
&&& \\
{\bf Magnitudes/fluxes} &&&& \\
$r^{1.2}$ & 16.29${\pm}$0.03 & 16.07${\pm}$0.03 & 14.29${\pm}$0.03 & 13.97${\pm}$0.03 \\
24$\mu$m$^3$ & 4007${\pm}$0.222\,mJy & 11.013${\pm}$0.573\,mJy & 42${\pm}$2\,mJy & \\
1.4GHz$^4$ & 0.89\,mJy & 2.30\,mJy & 9.31\,mJy \\
&&&& \\
{\bf Masses} &&&& \\
stellar mass$^{5,6}$ & $3.61\times10^{9}$M$_\odot$ & $1.0\times10^{10}$M$_\odot$ & $7.7\times10^{10}$M$_\odot$ & $1.6\times10^{11}$M$_\odot$ \\
total halo mass$^{5,6}$ & $1.9\times10^{11}$M$_\odot$ & $2.5\times10^{11}$M$_\odot$ &  $1.2.\times10^{12}$M$_\odot$ & \\
&&&& \\
{\bf Distances} &&&& \\
redshift$^{7}$ & 0.04261$\pm$0.00023 & 0.04817$\pm$0.00042 & 0.05152$\pm$0.00019 & 0.05277$\pm$0.00015 \\
projected distance$^{5,6}$ & 282\,kpc & 226\,kpc & 995\,kpc & 385\,kpc \\
to the parent cluster centre & $\sim 0.3r_{200}$ & $\sim 0.2r_{200}$ & $\sim 0.4r_{200}$ & $\sim 0.3r_{200}$ \\
&&&& \\
{\bf Star formation rates} &&&& \\
UV+IR global SFR$^{3}$ & 1.4\,M$_\odot$yr$^{-1}$ & 2.5\,M$_\odot$yr$^{-1}$ & 8.53\,M$_\odot$yr$^{-1}$ & \\
H$\alpha$ global SFR$^{5,6}$ & $1.76\pm 0.56$\,M$_\odot$yr$^{-1}$ & $3.49\pm 1.07$\,M$_\odot$yr$^{-1}$ & $7.2\pm 2.2$\,M$_\odot$yr$^{-1}$ & 6\,M$_\odot$yr$^{-1}$\\
&&&& \\
\hline
\end{tabular}}
\begin{tabular}{l}
Magnitudes are in the AB photometric system and corrected for galactic extinction.\\
  Sources: $^1$\citet{ShaSSI}; $^2$\citet{ShaSSII}; $^3$\citet{ACCESSII}; $^4$\citet{M05}; $^5$\citet{ACCESSV} for SOS\,114372; \\
  $^6$\citet{ShaSSIV} for SOS\,61086 and SOS\,90630;
  $^7$\citet{ShaSSV}. 
\end{tabular}
\label{gals}
\end{table*}


We report on four radio continuum tails (see Fig.~\ref{fig:gals})
associated to RPS galaxies which are supercluster members and
investigate if and to which extent ram-pressure affects the
non-thermal component of these galaxies and discuss our results in the
light of the ongoing dynamical activity in the Shapley
supercluster. The paper is organised as follows.  In
Section~\ref{targets} we provide a short description of the selection
and properties of the RPS galaxies in our sample. The radio
observations and data reduction are described in
Section~\ref{radioobs} where the radio properties of the galaxies are
also derived. In Section~\ref{RT}, complementing the radio
observations with multi-band data, we analyse the radio tails in the
framework of the RPS scenario.  We investigate the origin of the
electrons powering the radio tails and highlight the role of the
environment in Section~\ref{env} and in Section~\ref{sumcon} the main
results are summarised.

Throughout the paper we adopt a cosmology with $\Omega_M$=0.3,
$\Omega_\Lambda$= 0.7, and H$_0$=70\,km\,s$^{-1}$Mpc$^{-1}$. According
to this cosmology 1\,arcsec corresponds to 0.96\,kpc at the median
redshift ($z \sim 0.048$) of the Shapley supercluster. The radio
spectral energy distribution is modelled as a power law $S = A \nu^{\alpha}$.

\begin{table*}
  \centering
\caption{Logs of the radio observations}
{\small
\begin{tabular}{rcccrrcl}
\hline
    {\bf Frequency}  & {\bf Array} & {\bf Project ID} & {\bf Obs. date} & {\bf Bandwidth} & {\bf Time} & {\bf Resolution} & {\bf rms} \\
    MHz & & & & MHz & hr & arcsec$\times$arcsec & mJy/beam\\
    \hline
 235 & GMRT    & 30\_024   & 22-05-2016 &  32 &  7 & 24.45$\times$10.77 & 0.76  \\
 325 & GMRT    & 22\_039   & 30-06-2012 &  32 &  7 & 14.05$\times$ 9.53 & 0.13  \\
 400 & uGMRT   & 42\_019   & 06-06-2022 & 200 &  7 &  8.17$\times$ 4.78 & 0.028 \\
 400 & uGMRT   & 42\_019   & 10-06-2022 & 200 &  7 &  8.05$\times$ 4.51 & 0.028 \\
 610 & GMRT    & 22\_039   & 02-05-2015 &  32 &  5 &  9.54$\times$ 5.04 & 0.11  \\
 887 & ASKAP $^a$  & ESP~20    & 19-03-2019 & 288 & 11 & 13.23$\times$10.43 & 0.026 \\
 943 & ASKAP $^b$ & SB43137 & 03-08-2022 & 288 & 10 & 18.00$\times$18.00 & 0.025 \\
1283 & MeerKAT & MGCLS $^c$ & 06-08-2018 & 856 &  8 &  7.66$\times$ 7.28 & 0.007 \\   
1367 & ASKAP $^b$ & SB43206 & 06-08-2022 & 288 &  8 & 10.90$\times$ 7.80 & 0.033 \\
&&&&&& \\
\hline
\end{tabular}}
\begin{tabular}{l}
$a$: EMU Evolutionary Map of the Universe \citep{Norris11}, Early Science Project ESP\,20 \citep{Venturi+22}. \\
$b$: POSSUM \citep[Polarization Sky Survey of the Universe's Magnetism,][]{Gaensler+10} Pilot\,2 survey. \\
$c$: Pointing belonging to the MeerKAT Galaxy Cluster Legacy Survey \citep[MGCLS,][]{Knowles+22}.  \\
\end{tabular}
\label{radiodata}
\end{table*}

\section{RPS galaxies in the Shapley supercluster}
\label{targets}

The Shapley supercluster \citep{Shapley30} is the most massive
supercluster in the local Universe. It encompasses at least 25 Abell
clusters in the redshift range $0.035<z<0.06$ over a $\sim
15\times10$\,deg$^2$ region
\citep{R89,EFW97,DSE05,PQC06,Quintana+20}. At the heart of the Shapley
supercluster is the high-density Shapley Supercluster Core (SSC, at
median redshift $z=0.048$), comprising three Abell clusters and two
poor clusters, forming a continuous filamentary structure 2\,deg
($\sim7$\,Mpc) in extent filled with hot gas
\citep[][]{Planck13}. Evidence of cluster interaction and merging in
the SSC were extensively provided by dynamical studies
\citep[e.g.][]{Bardelli+98}, X-ray observations
\citep[e.g.][]{Finoguenov+04,Rossetti+07} and radio continuum analysis
\citep[e.g.][]{Venturi+00,Venturi+03,Giacintucci+05,Giacintucci+22}. Over
the past years, the Shapley Supercluster Survey
\citep[ShaSS][]{ShaSSI} traced the filaments feeding the SSC, and
revealed how the surrounding clusters are connected to each other and
to the SSC \citep{ShaSSI,ShaSSV,ShaSSVI}.

The four galaxies analysed in this work have been identified in the
region of the Shapley supercluster covered by the ShaSS by
\cite{ACCESSV,ShaSSIII}. From a spectroscopic catalogue, which is 95
per cent complete at magnitude $i < 18$, the RPS candidates were
selected in the deep (down to magnitude $r\sim 24$ at 5$\sigma$)
images collected at the ESO VLT Survey Telescope (VST) according the
criteria introduced by \cite{ESE14}: (i) disturbed morphology
indicative of a unilateral external force; (ii) brightness knots and
colour gradients suggesting bursts of SF; (iii) evidence of tails. The
final sample of RPS candidates turned out to constitute the 0.5 per
cent of the supercluster galaxies in agreement with the 0.6 per cent
found by \citet{ESE14}. Among the 13 selected galaxies 9 actually
presented extraplanar ionised gas and only 4 turned out to be
significantly affected by ram pressure. This does not imply that these
are the only galaxies affected by ram pressure across ShaSS, but that
the visual selection of the RPS candidates in the ShaSS imaging
allowed the identification of these cases \citep[cf.][]{BFS22}. All
four galaxies are members of the supercluster although associated with
different clusters.

Ongoing RPS has been already ascertained for three galaxies
(SOS\,114372, SOS\,90630, SOS\,61086) thanks to the observations
carried out with the Wide-Field Spectrograph
\citep[WiFeS][]{Dopita07,Dopita10}, then complemented by
$N$-body/hydrodynamical simulations \citep[see][]{ACCESSV,ShaSSIII}.
The fourth galaxy, ShaSS\,421015419 (hereafter ShaSS\,421) has been
observed with MUSE - Multi-Unit Spectroscopic Explorer
\citep{Bacon10}. The analysis of this particular object is in progress
and it will be published in detail elsewhere (Merluzzi et al., in
preparation), however there is evidence that this galaxy is also
affected by RPS as shown by the preliminary results presented
here. Therefore for sake of completeness, we include this object in
the sample.

In Table~\ref{gals} the main global properties of each galaxy are
listed. In Fig.~\ref{fig:gals} the positions of the four galaxies
across the ShaSS are indicated together with those of the clusters of
galaxies.

The radio tail of SOS\,61086 has been already studied in
\citet{Venturi+22}, however, in Section~\ref{SOS-61086} we recall the
main results for a more comprehensive discussion.

\begin{table*}
  \centering
\caption{Flux density values from images convolved to 18$\times$18\,arcsec$^2$.}
{\small
\begin{tabular}{lrlccc}
\hline
    {\bf Galaxy}  & {\bf Frequency} & {\bf Telescope Array} & {\bf S$_{\rm tot}$} & {\bf S$_{\rm gal+tail1}$}  & {\bf S$_{\rm tail2}$} \\
    & MHz & & mJy & mJy & mJy \\
    \hline

{\bf SOS\,90630} & 235 & GMRT & 9.01$\pm$ 0.90 &    $a$          &   $a$            \\
                 & 325 & GMRT & 6.92$\pm$ 0.63 &  5.90$\pm$ 0.59&   1.08$\pm$ 0.19\\
                 & 400 & uGMRT & 5.63$\pm$ 0.47 &  5.39$\pm$ 0.43&   0.27$\pm$ 0.07\\
                 & 887 & ASKAP $^b$ & 3.72$\pm$ 0.20 &  3.63$\pm$ 0.18&   0.22$\pm$ 0.03\\
                 & 943 & ASKAP $^c$ & 3.82$\pm$ 0.20 &  3.57$\pm$ 0.18&   0.18$\pm$ 0.03\\
                 & 1283 & MeerKAT & 2.89$\pm$ 0.12  &  2.68$\pm$ 0.10 &  0.12$\pm$ 0.01\\ 
                 & 1367 & ASKAP $^c$ & 2.52$\pm$ 0.15  &  2.46$\pm$ 0.13 &  0.12$\pm$ 0.04\\
 &&&&&\\
\hline
 {\bf SOS\,114372} &  235 & GMRT & 21.93$\pm$ 3.58 &     $a$            &    $a$       \\
                   &  325 & GMRT & 30.56$\pm$ 2.50 &  24.77$\pm$ 2.48  &  5.88$\pm$ 0.84\\
                   &  400 & uGMRT & 27.13$\pm$ 2.18 & 22.64$\pm$ 1.81  &  5.70$\pm$ 0.49\\ 
                   &  887 & ASKAP $^b$ & 16.70$\pm$ 0.84 &  14.39$\pm$ 1.44  &  2.86$\pm$ 0.16\\
                   &  943 & ASKAP $^c$ & 15.82$\pm$ 0.80 &  13.88$\pm$ 1.39  &  2.16$\pm$ 0.13\\
                   & 1283 & MeerKAT & 11.67$\pm$ 0.47 &  10.35$\pm$ 1.04  &  1.42$\pm$ 0.11\\
                   & 1367 & ASKAP $^c$ & 11.09$\pm$ 0.57 &   9.54$\pm$ 0.48  &  1.18$\pm$ 0.10\\
 &&&&&\\
\hline
\end{tabular}}
\begin{tabular}{l}
$a$: See Section~\ref{radioimag}. \\
$b$: EMU Evolutionary Map of the Universe \citep{Norris11}, Early Science Project ESP\,20 \citep{Venturi+22}. \\
$c$: POSSUM \citep[Polarization Sky Survey of the Universe's Magnetism,][]{Gaensler+10} Pilot\,2 survey. \\  
\end{tabular}
\label{radiofluxes18}
\end{table*}

\section{Radio data}
\label{radioobs}

The galaxies SOS\,114372, SOS\,90630 and SOS\,61086 are covered by
several radio observations of the SSC over a frequency range from
235\,MHz to 1367\,MHz, carried out with a variety of radio
interferometers. In particular, the data were collected with {\it i})
GMRT \citep{Swarup+91} and uGMRT \citep{Gupta+17} at 235\,MHz,
325\,MHz, 400\,MHz (Band\,3) and 610\,MHz; {\it ii}) MeerKAT
\citep{Camilo18} at 1.28\,GHz; and {\it iii}) the Australian Square
Kilometre Array Pathfinder \citep[ASKAP,][]{Hotan21} at 887\,MHz (EMU
Evolutionary Map of the Universe, \citet{Norris11}, Early Science
Project ESP\,20, \citet{Venturi+22}), 943\,MHz and 1367\,MHz as part
of the POSSUM \citep[Polarization Sky Survey of the Universe's
  Magnetism,][]{Gaensler+10} Pilot\,2 survey. On the other hand, for
the galaxy ShaSS\,421 only ASKAP imaging at 887\,MHz (Early Science
Project ESP\,20) is available at present. The summary of the
observations used in this paper is presented in
Table~\ref{radiodata}. For projects 22\_039, 30\_024 and 42\_019 the
parameters of the full resolution images are reported.

The observing details and data reduction strategy concerning projects
30\_024, 22\_039, ESP\,20 and MGCLS \citep[MeerKAT Galaxy Cluster
  Legacy Survey,][]{Knowles+22} were provided in \citet{Venturi+22}
where the radio tail of galaxy SOS\,61086 was also studied. MGCLS
images at full resolution (see Table~\ref{radiodata}) and at the
resolution of 15$^{\prime\prime}\times15^{\prime\prime}$ have been
used.  Beyond the total intensity images, the MGCLS data products
provide in-band spectral index (in the range 908-1656\,MHz) and full
Stokes information.

Project 42\_019 consists of 5 uGMRT pointings in Band\,3
(250-550\,MHz) which cover the SSC from west of the cluster A\,3558 to
east of A\,3562. It was designed to study the diffuse emission
features in this region (Trehaeven et al. in prep). The two datasets
which include SOS\,90630 and SOS\,114372 are centred at RA$_{J2000}$ =
$13^h30^m0^s$ and DEC$_{J2000}$ =
$-31^{\circ}39^{\prime}00^{\prime\prime}$ and RA$_{J2000}$ =
$13^h27^m58^s$ and DEC$_{J2000}$ =
$-31^{\circ}30^{\prime}30^{\prime\prime}$ for the two sources
respectively.  The data were recorded with 8192 spectral channels and
an integration time of 4\,sec in full Stokes mode. The sources 3C\,48
and 3C\,236 were used as primary calibrators, and 1311--222 was used
as phase calibrator. The data were processed by means of SPAM
\citep[Source Peeling and Atmospheric Modelling,][]{Intema+09}. The
full band was split into 6 sub-bands each 33.3\,MHz wide and the
sub-band were then imaged together using WSClean (v2.10) at the common
frequency of 400\,MHz. Residual amplitude errors are of the order of
8\%.  The final deep images were produced with {\it
  weighting=`Briggs'} using {\it robust=-0.5} and a set of tapering to
properly image the features on different angular scales. Primary beam
correction was performed using the task PBCOR in the NRAO Astronomical
Image Processing System (AIPS).

All galaxies discussed in this paper (SOS\,61086, SOS\,90630,
SOS\,114372 and ShaSS\,421) are included in the $\sim 6\times$6
deg$^2$ field imaged by ASKAP at 887\,MHz as part of the EMU Early
Science Programme \citep{Johnston08,Norris11} ESP\,20 and already
presented in \citet{Venturi+22}, where observation and data reduction
details are given.  In brief, the telescope was configured to produce
36 electronically formed beams arranged on the sky in a 6$\times$6
square grid; each individual beam covers an area of $\sim 1$\,deg$^2$
for a total instantaneous field of view of $\sim 31$\,deg$^2$.  The
observations were performed with 35 out of the 36 antennas of the
array, at 887\,MHz, with bandwidth $\Delta\nu$=288\,MHz (288 channels
each of bandwidth of 1\,MHz). The final image is centred at RA$_{\rm
  J2000}=13^h25^m50^s$, DEC$_{\rm
  J2000}=-31^{\circ}03^{\prime}05^{\prime\prime}$.  Residual amplitude
calibration errors are of the order of 5\%.

Finally, the galaxies SOS\,61086, SOS\,90630, SOS\,114372 were
observed by ASKAP as part of the POSSUM Pilot 2 survey (POSSUM2),
whose goal is to study the magnetic fields of the Universe in
different environments.  The pointing centres for Band\,1
(800-1088\,MHz) and Band\,2 (1152-1440\,MHz) are respectively
RA$_{J2000}$ = $13^h29^m47^s$ and DEC$_{J2000}$ =
$-30^{\circ}17^{\prime}10^{\prime\prime}$ and RA$_{J2000}$ =
$13^h28^m51^s$ and DEC$_{J2000}$ =
$-29^{\circ}43^{\prime}04^{\prime\prime}$. The Band\,1 data was taken
in the {\it closepack36} PAF beam configuration and Band\,2 was in the
square\_6$\times$6 configuration. Due to the different frequencies,
the final image in Band\,2 covers a slightly smaller sky area, each of
the electronically formed beam being $\sim 30$\% smaller than in
Band\,1. The data reduction of the POSSUM2 observations is the same as
the procedure followed for the Rapid ASKAP Continuum Survey (RACS),
detailed in \citet{Duchesne+23}. Residual amplitude calibration errors
are of the order of 5\%. The datasets in both bands provide total
intensity images, polarisation information through Rotation Measure
(RM) synthesis, and spectral index information (obtained dividing the
Taylor 1 by the Taylor 0 images). The final full resolution images of
the galaxies SOS\,90630 and SOS\,114372 at all radio frequencies are
reported in the Appendix~\ref{AppA}. The analysis of the polarisation
will be presented in a forthcoming paper.

\begin{figure}
\begin{centering}
\includegraphics[width=85mm]{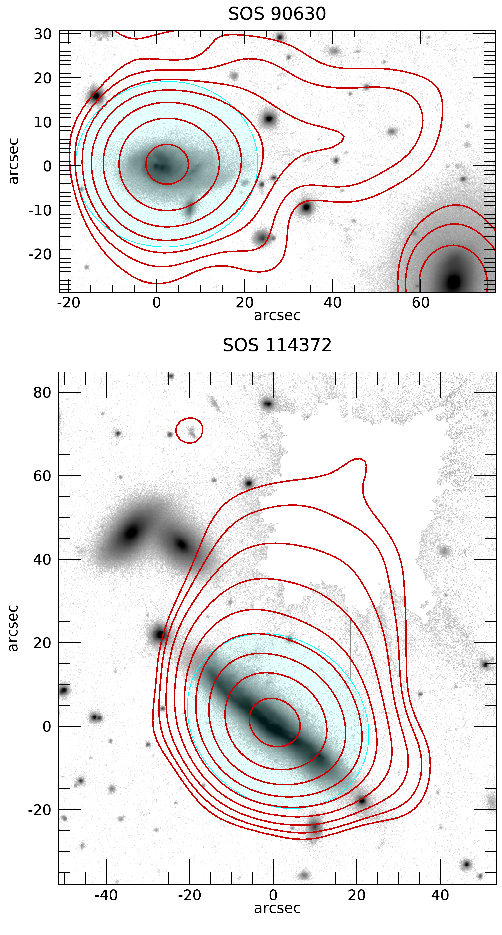}
\end{centering}
\caption{VST $r$-band images with superimposed the 1.28\,GHz MeerKAT
  isophotes convolved to 18$\times$18\,arcsec$^2$ of SOS\,90630 (upper
  panel) and SOS\,114372 (lower panel). The cyan regions denote the
  fitted Gaussian component (see text) up to r=1$\times$FWHM. The
  radio contours are spaced by a factor 2 in flux.
\label{fig:masks}}
\end{figure}

\subsection{Radio images and fluxes}
\label{radioimag}

Our aim is to investigate the nature of the extraplanar radio
continuum emission, i.e. to identify the mechanisms responsible of
such a feature as well as which processes are powering the electrons
outside the galaxy disk. The multi-band radio data with their
resolution and sensitivity enable us to derive the radio properties in
different regions - discs and tails. However, the different angular
resolutions in our dataset (see Table~\ref{radiodata}) need to be
considered in defining the galaxy regions. For SOS\,90630 and
SOS\,114372, we can distinguish two regions: one comprehending the
galaxy and the inner part of the tail (hereafter `gal+tail1') and the
other covering the outer part of the tail (tail2). This choice was
made to account for the different angular resolutions of our radio
images.

For SOS\,90630 in Fig.~\ref{fig:x8_090630} we notice that in the full
resolution images the length of the tail is similar ($\sim 43$\,kpc)
at four frequencies, while at 235, 610 and 1367\,MHz the sensitivities
do not (or only barely) allow the detection of the radio tail outside
the visible galaxy (see Table~\ref{radiodata}).

In Fig.~\ref{fig:x8_114372} the tail of SOS\,114372 shows, at the
sensitivity of our images, different extensions in projection at
different frequencies, e.g. $\sim 72$\,kpc at 943\,MHz and 29\,kpc at
1367\,MHz. We notice that the ASKAP-POSSUM2 image at 943\,MHz has the
second better sensitivity in the dataset and traces the extraplanar
emission further out the galaxy disk.

To account for the different angular resolutions in our datasets,
which range from $7.66^{\prime\prime}\times7.28^{\prime\prime}$ of the
full-resolution MeerKAT images to
$18^{\prime\prime}\times18^{\prime\prime}$ for ASKAP-POSSUM Band 1,
and ensure a consistent spectral analysis, we convolved all images to
the same angular resolution of
$18^{\prime\prime}\times18^{\prime\prime}$, with the exception of GMRT
235\,MHz for SOS\,90630 (whose beam is larger) and GMRT 610\,MHz
(which is not used in the analysis, see below). The convolved images
are shown in Figs.~\ref{fig:x3_090630} and \ref{fig:x3_114372}.

At each frequency, for SOS\,90630 and SOS\,114372 we derived {\it i})
the integrated flux density; {\it ii}) the flux density of gal+tail1;
{\it iii}) the flux density of the remaining outer part of the tail
(tail2). At the adopted resolution the gal+tail1 component is all
included within the beam of the image, and S$_{\rm gal+tail1}$ has
been obtained fitting the peak in the image with a Gaussian component
(see Fig.~\ref{fig:masks}). The total flux, S$_{\rm tot}$, was derived
by integration over the whole area taking the images at 943\,MHz as
reference. The flux of tail2 is obtained subtracting S$_{\rm
  gal+tail1}$ from S$_{\rm tot}$.  These values are listed in
Table~\ref{radiofluxes18}.

For each galaxy and each component we estimated the uncertainty
$\Delta S$ on the flux density $S$ using the relation

  \begin{equation}
  $$\Delta S=\sqrt{(\sigma\times\sqrt{N_{\rm beam}})^2+(\xi_{\rm cal}\times S)^2}$$
\end{equation}

\noindent  
where $\sigma$ is the local noise in the image, $N_{\rm beam}$ is the
extent of the source (or component) in number of beams, $\xi_{\rm
  cal}$ is the calibration uncertainty.

For ShaSS\,421 we derived S$_{\rm tot}$=5.16$\pm$0.26\,mJy.  All the
other flux density values are reported in Table~\ref{radiofluxes18}.

In the following we provide a description of the radio morphology for
each galaxy and a zero-order analysis of the integrated spectrum of
the region encompassing the galaxy and the beginning of the radio tail
(gal+tail1), and of the outer tail region (tail2). In the description
of the morphologyy we benefit of the high-resolution MeerKAT images,
although the quantitative analysis is based on the convolved
images.

\begin{figure}
\begin{centering}
\includegraphics[width=85mm]{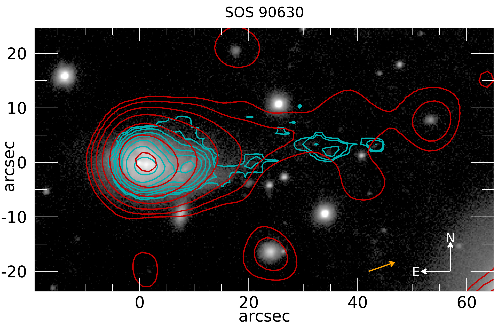}  
\end{centering}
\caption{VST $r$-band image of SOS\,90639 ($\sim 0.6^{\prime\prime}$
  angular resolution) with superimposed the 1.28\,GHz MeerKAT radio
  contours (red, $7^{\prime\prime}$ angular resolution) and the
  H$\alpha$ contours (cyan, $\sim1.5^{\prime\prime}$ angular
  resolution). The radio outer isophote corresponds to 3$\sigma$
  (0.021\,mJy\,b$^{-1}$), the subsequent isophotes are spaced by a
  factor 2 in flux. For the H$\alpha$ data from WiFeS, the outer
  isophote corresponds to
  5$\times$10$^{-18}$\,erg\,s$^{-1}$\,arcsec$^{-2}$ (5$\sigma$ in
  WiFeS spectra), the other isophotes are spaced by a factor 2 in
  flux. Bottom right: the orientation of the field and the direction
  of the wind velocity (orange arrow, see text) are indicated.
\label{fig:90630_r-band_radio_Ha}}
\end{figure}

\bigskip
{\it SOS\,90630}. This galaxy shows a radio continuum tail oriented
westwards with respect to the position of the optical galaxy centre,
coincident with the radio peak. At 1.28\,GHz it extends for 45\,arcsec
($\sim43$\,kpc) with hints that the radio tail forks in two branches
perpendicular to the direction of the tail at 40\,arcsec west from the
centre, as seen in the MeerKAT 1283\,MHz image at full resolution
(Fig.~\ref{fig:90630_r-band_radio_Ha}). We point out that the
westernmost knot is most likely associated with the background galaxy
visible in Fig.~\ref{fig:90630_r-band_radio_Ha} and for this reason we
do not consider it as part of the tail. The total radio power of
SOS\,90630 at 1367\,MHz is P$_{\rm 1367\,MHz}=1.49\times10^{22}$
W\,Hz$^{-1}$.

In order to check for spectral steepening along the tail, we separated
the flux density of the radio emission in the two components as
explained above.  Usings the flux density obtained from the
$18^{\prime\prime}\times 18^{\prime\prime}$ images for each component
we fitted the spectra of these two components with a power law (the
610 MHz flux density values were not included in the fit as they are
clearly overestimated). Despite the scatter of our measurements along
the tail, mainly due to the different radio interferometers used in
this analysis, our data show an unambiguous steepening, with
$\alpha_{\rm 325\,MHz}^{\rm 1367\,MHz}$(gal+tail1)=$-0.63\pm0.11$ and
$\alpha_{\rm 325\,MHz}^{\rm 1367\,MHz}$(tail2)=$-1.03\pm0.12$. The
integrated spectrum of the radio emission has a spectral index
$\alpha_{\rm 235\,MHz}^{\rm 1367\,MHz}$(tot)=$-0.62\pm0.09$.

\begin{figure}
\begin{centering}
\includegraphics[width=85mm]{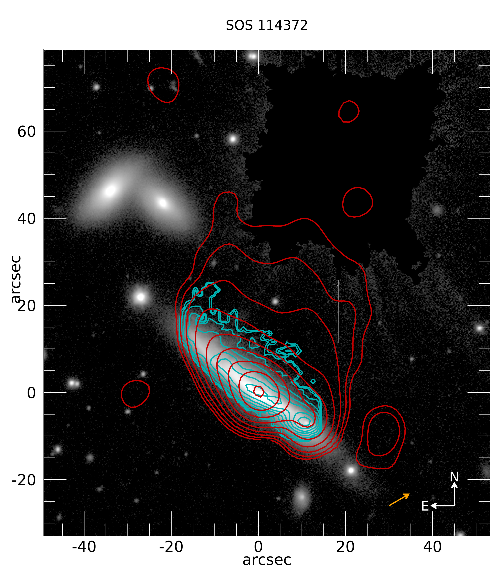}  
\end{centering}
\caption{VST $r$-band image of SOS\,114372 ($\sim 0.6^{\prime\prime}$
  angular resolution) with superimposed the 1.28\,GHz MeerKAT radio
  contours (red, $7^{\prime\prime}$ angular resolution) and the
  H$\alpha$ contours (cyan, $\sim1.5^{\prime\prime}$ angular
  resolution). The contours are drawn as in
  Fig.~\ref{fig:90630_r-band_radio_Ha}. Botton right: the orientation
  of the field and the direction of the wind velocity (orange arrow,
  see text) are indicated.
\label{fig:114372_r-band_radio_Ha}}
\end{figure}

\bigskip
{\it SOS\,114372}. The radio emission in this galaxy follows the
optical disc (but see Section~\ref{RT_114372}), for a total size of
$\sim 30$\,kpc, and is characterised by a radio tail roughly
perpendicular in projection to the optical disc in the north-west
direction, as shown in Fig.~\ref{fig:114372_r-band_radio_Ha}. The
radio tail is resolved along its minor axis, and shows two ridges of
emission extending $\sim40$\,arcsec ($\sim38$\,kpc) north of the disc,
clearly visible at all frequencies in the higher resolution
images. The total radio power at 1367\,MHz for this source is P$_{\rm
  1367\,MHz}=6.35\times10^{22}$ W\,Hz$^{-1}$.

As in the case of SOS\,90630, using the flux density values derived
from the $18^{\prime\prime}\times18^{\prime\prime}$ images a single
power law fit shows a clear spectral steepening in tail2 compared to
the combination of the galaxy and beginning of the tail.  In
particular, we measured $\alpha_{\rm 325\,MHz}^{\rm
  1367\,MHz}$(galaxy+tail1)=$-0.72\pm0.10$ and $\alpha_{\rm
  325\,MHz}^{\rm 1367\,MHz}$(tail2)=$-1.49\pm0.12$. The integrated
spectrum of the radio emission is dominated by the galaxy and has a
spectral index $\alpha_{\rm 325\,MHz}^{\rm
  1367\,MHz}$(tot)=$-0.70\pm0.10$.

\begin{figure}
\begin{center}
\includegraphics[width=80mm]{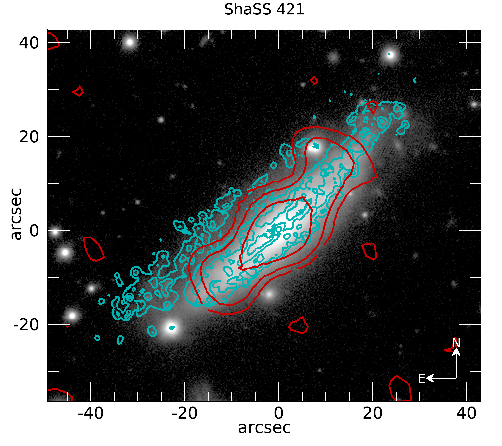}   
\end{center}
\caption{VST $r$-band image of ShaSS\,421 ($\sim 0.6^{\prime\prime}$
  angular resolution) with superimposed the ASKAP contours (red,
  restoring beam $13.2^{\prime\prime} \times 10.4^{\prime\prime}$).
  The radio outer isophote corresponds to 3$\sigma$
  (0.15\,mJy\,b$^{-1}$, here we adopted the local noise since the
  galaxy is close to the image border), the subsequent isophotes are
  spaced by a factor 2 in flux. For the H$\alpha$ data from MUSE
  (cyan), the outer isophote corresponds to
  1$\times$10$^{-18}$\,erg\,s$^{-1}$\,arcsec$^{-2}$, the other
  isophotes are spaced by a factor 4 in flux.}
\label{fig:ShaSS421_Ha_ASKAP}
\end{figure}

\bigskip
{\it ShaSS\,421}. Only the 887\,MHz ASKAP observation is available for
ShaSS\,421. No radio tail is detected at the sensitivity of the image,
whose local noise is $\sim 50$\,$\mu$Jy\,b$^{-1}$, significantly
higher than for SOS\,90630 and SOS\,114372. The radio emission extends
for $\sim 62$\,kpc along the optical disc, as shown in
Fig.~\ref{fig:ShaSS421_Ha_ASKAP}. The radio morphology has a bipolar
shape. The radio continuum emission detected by ASKAP peaks in the
centre of the galaxy stellar disc. Moving away from the centre along
the galaxy major axes in both sides of the disc, the radio emission
shows recollimation, to broaden again in a `wrapped candy'
morphology. The radio power of the emission is P$_{\rm
  887\,MHz}=3.45\times10^{22}$ W\,Hz$^{-1}$. The radio morphology is
actually suggestive of a radio galaxy of nuclear origin. We extracted
the 1.4\,GHz flux density from the NRAO VLA Sky Survey
\citep[NVSS,][]{Condon+98} and obtained a total spectral index
$\alpha_{\rm 887\,MHz}^{\rm 1400\,MHz}=-1.04\pm0.18$. Further studies
on the origin of the radio emission are ongoing.

\section{Radio tails of RPS galaxies}
\label{RT}

\begin{figure}
  \begin{centering}
\includegraphics[width=85mm]{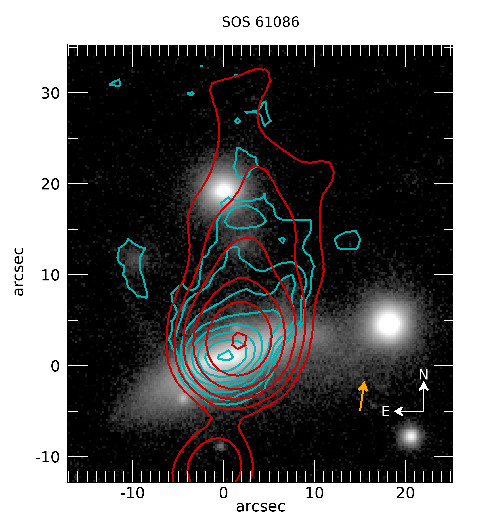}
  \end{centering}
  \caption{VST $r$-band image of SOS\,61086 ($\sim0.6^{\prime\prime}$
    angular resolution) with superimposed the 1.28\,GHz MeerKAT radio
    contours (red, $7^{\prime\prime}$ angular resolution) and the
    H$\alpha$ contours (cyan, $\sim 1.5^{\prime\prime}$ angular
    resolution). The contours are drawn as in
    Fig.~\ref{fig:90630_r-band_radio_Ha}. Botton right: the
    orientation of the field and the direction of the wind velocity
    (orange arrow, see text) are indicated.
\label{fig:61086_r-band_radio_Ha}}
\end{figure}

In this section we analyse the observed radio properties of
SOS\,90630, SOS\,114372 and ShaSS\,421 in the framework of the RPS
scenario as supported for these objects by multi-band studies and
hydrodynamical simulations. We also recall the main results obtained
by \citet{Venturi+22} for SOS\,61086.  In
Figs.~\ref{fig:90630_r-band_radio_Ha},
\ref{fig:114372_r-band_radio_Ha}, \ref{fig:ShaSS421_Ha_ASKAP} and
\ref{fig:61086_r-band_radio_Ha} we show the VST $r$-band image of each
galaxy with superimposed the radio continuum emission at 1.28\,MHz
(red contours) and the H$\alpha$ emission (cyan contours).

\subsection{SOS\,90630}
\label{RT_90630}

The galaxy SOS\,90630 is a member of the poor cluster
SC\,1327-312. The VST $r$-band image
Fig.~\ref{fig:90630_r-band_radio_Ha} shows the highly asymmetric light
distribution with a western arm extending well out of the disc. The
deep MeerKAT radio continuum emission at 1.28\,MHz peaks at the galaxy
centre and extends west in the same direction of the asymmetric
emission observed in the visible image.

To estimate the equipartion magnetic field for the galaxy and inner
tail (gal+tail1) and for the outer part of the tail (tail2) we used
the formulae in \citet{GF04}. For both regions we assumed a
cylindrical volume, a ratio between electrons and protons k=1 and a
filling factor $\Phi$=1, obtaining H$_{\rm eq}\sim 1.8$\,$\mu$G and
$\sim 0.9$\,$\mu$G, respectively.

In order to estimate the age of the radiating electrons in the two
regions under considerations, we derived the break frequency $\nu_{\rm
  br}$ by fitting the synchrotron spectrum with the software Synage
\citep{Murgia+99}. The best fit is obtained assuming a JP model
\citep{JP73} and an injection spectral index $\alpha_{\rm
  inj}=-0.5$. The fit provides a break frequency $\nu_{\rm br}$=
10327\,MHz for gal+tail1, and $\nu_{\rm br}$=1994\,MHz for tail2.  The
fit of the radio emissions are shown in Fig.~\ref{fig:spectralfits},
upper row.  The corresponding radiative ages, computed following Eq.~1
in \citet{Feretti+98} which accounts for the equivalent magnetic field
strength of the CMB radiation, are t$_{\rm rad}\sim 4.0\times10^7$ yr
for the galaxy and inner tail and t$_{\rm rad}\sim 7.7\times10^7$ yr
for the outer tail.

\begin{figure}
\begin{centering}
\includegraphics[width=85mm]{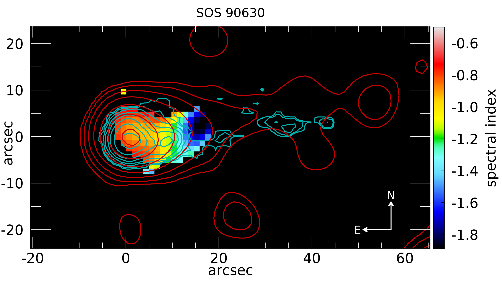} 
\end{centering}
\caption{The MeerKAT inband spectral index in the frequency range
  908-1656\,MHz. Red and cyan contours denote the radio continuum and
  H$\alpha$ emission, respectively. The MeerKAT spatial resolution is
  $7.66^{\prime\prime} \times 7.28^{\prime\prime}$ (p.a. 68.17). The
  contours are drawn as in Fig.~\ref{fig:90630_r-band_radio_Ha}.
\label{fig:90630_index}}
\end{figure}

The MeerKAT inband spectral index in the frequency range 908-1656\,MHz
could be derived only for the nuclear component and inner part of the
tail.  The surface brightness in the region corresponding to tail2 is
too low and falls below the sensitivity limit of each frequency
channel used to derive the inband spectral
index. Fig.~\ref{fig:90630_index} clearly shows the steepening from
the nuclear region, where the spectral index is $\alpha\sim -0.8$, out
to the edge of tail1 where $\alpha\sim -1.8$.

In Fig.~\ref{fig:90630_r-band_radio_Ha} we also show the distribution
of the H$\alpha$ emission from WiFeS IFS (cyan contours) as analysed
by \citet{ShaSSIII}. The gas out of the disc extends $\sim 4$\,kpc in
projection in the NW side, and along a more than 40\,kpc-long tail to
the west. The radio tail extends in projection with almost the same
length. The ionised gas disc is also truncated in the South-East
side. Due to the different resolution it is not clear if the
non-thermal emission of the galaxy is characterised by a similar
truncation.  Moreover, the moderate disc truncation observed in the
ionised gas distribution could be related to the galaxy inclination
angle whose impact on the amount of gas stripped from a disc is
significant when the wind is close to edge-on as was shown with
hydrodynamical and numerical simulations
\citep[e.g.][]{RB06,Jachym+09}. In this case much less gas is removed
from the galaxy reducing the truncation effect.

\citet{ShaSSIII} found that the star formation across SOS\,90630 is
highly asymmetric with a `crown' of H{\sc ii} regions tracing the
eastern disc and other star-forming regions in the western disc
elongated toward the tail in a spiral arm \citep[see their Fig.~17
  and][]{AK14}.

A single burst of star formation over the last 200\,Myr that is still
ongoing is responsible of the young stellar population detected in the
centre and in the eastern edge of the galaxy disc, which is likely
induced by the compression exerted by the ram pressure. Due to their
ageing as traced by the inband spectral index in
Fig.~\ref{fig:90630_index}, the electrons in the inner tail can be
originated by this ongoing star formation.

$N$-body/hydrodynamical simulations \citep[see Appendix~\ref{AppC}
  and][for details]{ShaSSIII} enabled to estimate the parameters of
the hydrodynamical interaction between the gas disc of the galaxy and
the ICM of host cluster SC1327-312 indicating a ram pressure acting
close to edge-on, a wind velocity V$_{ICM}$= 500\, km\,s$^{-1}$ and
the time of the ram-pressure onset of about 120\,Myr.

Together with the similar length of the radio tail and the ionised gas
tail, the consistency among {\it i}) the epoch of the onset of
stripping (120\,Myr); {\it ii}) the age of the youngest stellar
population ($<$200\,Myr) and {\it iii}) the age of the radio tail
($<$100\,Myr) supports a scenario where the radio emitting plasma and
the warm gas tails are related to the same event, i.e. both affected
by the ram pressure with similar efficiency \citep[cf. the Virgo
  galaxy NGC\,4438,][]{Vollmer+09}.

\subsection{SOS\,114372}
\label{RT_114372}

The bright star-forming spiral galaxy SOS\,114372 is the only galaxy
of the sample which is member of a rich cluster, A\,3558.  Extraplanar
ionised gas has been detected along the full extent of the galaxy disc
out to 13\,kpc in projection from it (cyan contours in
Fig.~\ref{fig:114372_r-band_radio_Ha}). Running
$N$-body/hydrodynamical simulations of RPS for this galaxy (see
Appendix~\ref{AppC}), \citet{ACCESSV} reproduced the observed
distribution and surface brightness of the ionised gas. In this case,
the wind angle turns out to be intermediate between face-on and
edge-on, the estimated wind velocity is $\sim 1400$\,km\,s$^{-1}$ and
the derived time of onset of the RPS is $t \sim 60$\,Myr.

MeerKAT observations reveal a widespread emission out of the galaxy
disc whose extension in projection largely overcomes (38\,kpc long at
1.28\,MHz) the ionised gas tail. In this case the truncated ionised
gas disc observed in the NE edge of the galaxy corresponds to a clear
truncation also in the radio continuum (see
Fig. ~\ref{fig:114372_r-band_radio_Ha}). We point out that the IFS
observations do not cover the whole galaxy, for the SW edge of the
galaxy IFS data are not available \citep[see the sharp end of the cyan
  contours line in Fig.~\ref{fig:114372_r-band_radio_Ha} and Fig.~1
  of][]{ACCESSV}.  Therefore, it is misleading to compare radio and
IFS observations in this part of the galaxy.

\begin{figure}
\begin{centering}
\includegraphics[width=85mm]{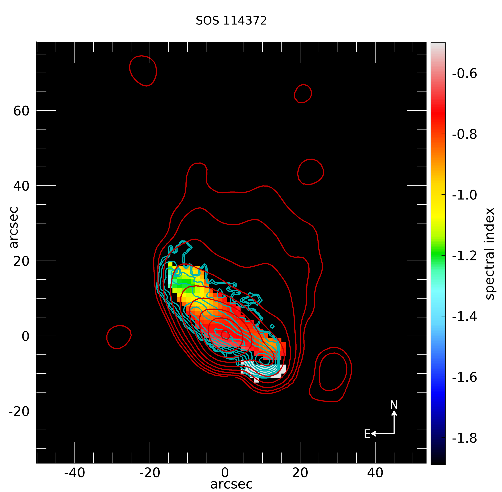}
\end{centering}
\caption{The MeerKAT inband spectral index of SOS\,114372 in the
  frequency range 908-1656\,MHz. Red and cyan contours denote the
  radio continuum and H$\alpha$ emission, respectively. The MeerKAT
  spatial resolution is $7.66^{\prime\prime} \times
  7.06^{\prime\prime}$ (p.a. 1.21). The contours are drawn as in
  Fig.~\ref{fig:90630_r-band_radio_Ha}.
\label{fig:114372_Ha_radio}}
\end{figure}

We computed the equipartition magnetic field and estimated the
radiative age of the electrons in the gal+tail1 and in the outer part
of the tail (tail2) under the same assumptions used for SOS\,90630. We
obtained H$_{\rm eq}\sim 2.56$\,$\mu$G and 0.86\,$\mu$G
respectively. We fitted the spectra of the two regions as for
SOS\,90630. Again the best fit was provided by a JP model (see
Fig.~\ref{fig:spectralfits}, lower row). We derived $\nu_{\rm
  br}=6851$\,MHz and $\nu_{\rm br}=1428$\,MHz for gal+tail1 and tail2,
respectively, with a corresponding age for the radiating electrons of
$t_{\rm rad}\sim 4.8\times10^7$\,yr and $t_{\rm rad}\sim
9.3\times10^7$\,yr.

The MeerKAT inband spectral index, shown in
Fig.~\ref{fig:114372_Ha_radio} with radio continuum and H${\alpha}$
contours overlaid, covers only the disc of the galaxy. It is flat in
the internal disc and consistent with the estimate of the spectral
index of the integrated spectrum throughout the whole extent of the
disc. We notice, however a marginal steepening at the north-eastern
edge of the disc and a region characterised by the higher values of
the spectral index SW in the disc. Unfortunately, due to sensitivity
limits, the inband spectral index along the tail could not be derived.

In a region SW from the centre \citet{ACCESSV} identified a
starburst. The agreement between the age ($\sim100$\,Myr) of the
youngest stellar population in this region and the age (started $\sim
60$\,Myr ago) of the gas stripping process inferred by the
$N$-body/hydrodynamical simulations led to ascribe this starburst to
the compression of the ISM caused by the ram pressure - one of the
first detections of ram-pressure induced star formation. On the other
hand, in the NE disc a region of star formation significantly lower
than in the rest of the disc was also identified. The full spectral
modelling and the line strengths provided evidence for the detection
of a starburst which lasted for $\sim 0.3$\,Gyr and which has been
just shut down. It is remarkable that these two regions correspond to
the regions where the extreme values of the radio spectral index are
measured in the disc as shown in Fig.~\ref{fig:114372_Ha_radio},
i.e. the spectral index steepens in the post-starburst NE region and
flattens in the SW starburst region of the disc. This shows how the
1.28\,MHz MeerKAT observations are able to resolve the star formation
in the Shapley Supercluster galaxies.

The flatness of the spectral index towards the SW starburst region is
consistent with the decreasing trend of the spectral index with
increasing SFR found in nearby galaxies covering a wide range in
morphology, SFR, stellar mass, and environment (field vs. interacting
galaxies) studied with both global \citep[e.g.][]{Tabatabaei+17} and
spatially resolved measurements \citep[e.g.][]{Westcott+18}. This
suggests that the star-formation process could affect the energetics
of the cosmic-ray electron population in a galaxy by increasing the
number of younger and more energetic relativistic particles in galaxy
regions with higher star-formation activity (e.g. starburst).

Also for this object the consistency among {\it i}) the epoch of the
onset of stripping ($\sim 60$\,Myr); {\it ii}) the age of the youngest
stellar population ($<$100\,Myr) and {\it iii}) the age of the radio
tail ($<$100\,Myr) suggest that the relativistic electrons and the
ionised gas are both stripped from the disc by the same
mechanism. However, unlike SOS\,90630, the length of the radio and
optical tails are significantly different.

\begin{figure}
  \begin{center}
\includegraphics[width=80mm]{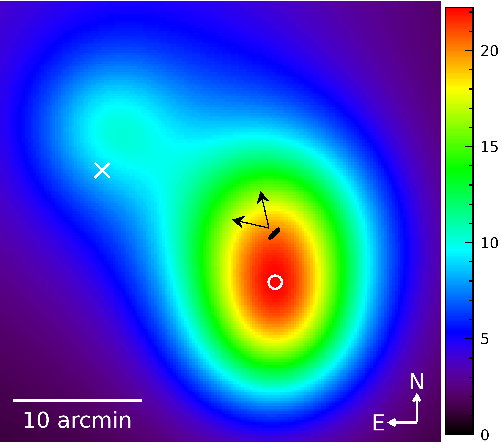}
\end{center}
\caption{The two-dimensional distribution of SSC galaxies in the
  region of AS\,0726 \citep[from][]{ShaSSV}. The small circle indicate
  the peak of the galaxy density. The small black ellipse north from
  this peak shows the galaxy (on scale) with the arrows indicating the
  range of directions of the gas outflow. The cross is the ROSAT
  X-centre by \citet{DSE05}. The density scale is in galaxies per
  Mpc$^2$.}
\label{fig:AS0726_gals_den}
\end{figure}

\subsection{ShaSS\,421}

ShaSS\,421 is the brightest and most massive galaxy of the sample. A
detailed study of this object, based on MUSE observations, is in
preparation. Here we briefly mention the main features supporting the
ongoing RPS scenario.

Fig.~\ref{fig:ShaSS421_Ha_ASKAP} shows evidence of extraplanar ionised
gas extending NE from the disc. Within a radius $\sim 250$\,kpc there
are not other cluster members with signs of disturbed morphology,
therefore excluding galaxy interaction as the cause of the observed
extraplanar emission. Further evidence in favour of the RPS scenario
is given by the orientation of the ShaSS\,421 tail which is opposite
to the peak of galaxy density as derived in the ShaSS
\citep{ShaSSI,ShaSSV}. The galaxy is located $\sim 385$\,kpc in
projection (0.3$r_{200}$) from the centre of AS\,0726 as shown in
Fig.~\ref{fig:AS0726_gals_den}.

The galaxy was targeted with MUSE. Two contiguous pointings slightly
overlapped in the centre covered an area of $\sim 1^\prime \times
2^\prime$ including the galaxy and its extraplanar region. In
Fig.~\ref{fig:ShaSS421_Ha_ASKAP} the ionised gas emission traced by
H$\alpha$ is shown. The extraplanar gas extends NW up to 15\,kpc out
of the disc and along all the disc length. The one-sided extraplanar
gas emission as well as the truncation of the gas disc along the SW
edge of the galaxy support an ongoing RPS event.

In this galaxy, we do not observe a radio tail.  The analysis of the
MUSE spectra showed that the galaxy host an AGN which could be
responsible of the radio emission intensity and shape. However, the
extent of the H$_{\alpha}$ emission is larger than the radio emission,
which does not reach either the outer edges of the galaxy or the
south-eastern extent above the galaxy disc. It seems that the RPS
mechanism is not capable to affect the non-thermal component of this
galaxy.

\subsection{SOS\,61086}
\label{SOS-61086}

The radio tail of SOS\,61086 has been studied in
\citet{Venturi+22}. Here we recall the main features of this object
and the proposed interpretation in order to compare them with those of
the other galaxies in the sample.

SOS\,61086 is a member of the poor cluster
SC\,1329-313. \citet{ShaSSIV} showed that this galaxy is characterised
by a tail of ionised gas extending up to $\sim 30$\,kpc in
projection. They demonstrated that the galaxy is affected by RPS and
estimated the time of the onset (`age') of RPS of about 250\,Myr in
agreement with the age of the young stellar population ($< 200$\,Myr),
which likely suggests a ram-pressure induced star formation.

\citet{Venturi+22} reported on the radio tail revealed at 1.28\,MHz
with MeerKAT. The radio-emitting plasma and the ionised gas show equal
extent in projection from the galaxy disc and similar truncation in
the gas disc itself (see Fig.~\ref{fig:61086_r-band_radio_Ha}). Taking
into account the steepening of the radio spectral index along the
tail, the age of the relativistic electrons and their velocity, which
is found to be consistent with the wind velocity inferred from the
$N$-body/hydrodynamical simulation, \citet{Venturi+22} concluded that
the radio tail revealed by MeerKAT observations in SOS\,61086 is due
to the ongoing RPS and it is fuelled by relativistic electrons
stripped from the galaxy.

\section{Discussion}
\label{env}

In Section~\ref{RT}, we provide evidence that the ionised gas tail and
that traced by the relativistic electrons are due to the same
mechanism - the hydrodynamical interaction between the ICM and the
ISM. A step further is to understand if the radio continuum emission
in the tail is powered either by {\it in situ} star formation
\citep[][]{Gavazzi+95} or by relativistic electrons stripped from the
galaxy by ram pressure and possibly re-accelerated
\citep[][]{Chen+20}.

Among the four RPS galaxies showing ionised gas tails, three present
consistent synchrotron tails revealed by multi-frequency radio
observations, while the 887\,MHz ASKAP observations do not detect
radio continuum extraplanar emission in the most massive galaxy of the
sample, ShaSS\,421. Beyond the different instrumental characteristics
(sensitivity and spatial resolution), which are crucial for the
detection of the faint extraplanar emission, the galaxy properties and
orbits as well as their own environment may play an important role in
the process of sweeping away the ISM from the disc.

In the following we investigate the origin of the electrons in the
tail and describe the different environments of the RPS galaxies in
our sample also in order to possibly distinguish the effects of the
magnetic field in the RPS process.

\subsection{Origin of the relativistic electrons in the tail of SOS\,90630}

The relative lengths of the radio continuum emission and the H$\alpha$
tails can provide a piece of information to disentangle the origin of
the relativistic electrons in the tail. Radio tails either longer
\citep[e.g.][]{Gavazzi+95} or shorter \citep[e.g.][]{Chen+20} than
H$\alpha$ ones have been observed. However, it is difficult to compare
the lengths of the tails because differences in extent may be due on
differences in sensitivity and projection effects.

The velocity of the plasma containing the relativistic electrons
stripped from the galaxy can be estimated from the tail length and the
radiative age. The age estimated for the relativistic electrons ($\sim
77$\,Myr) in tail2 is actually an average value throughout the extent
of the tail which can be associated to the average distance from the
centre of the outer tail, i.e. 30\,kpc (see Fig.~\ref{fig:masks}). The
tail length should be then corrected for the projection effect using
the angle between the ram-pressure wind velocity, as derived by the
simulations, and the line of sight. In this case this angle is $\phi
\sim 67$\,deg.

Under these assumptions, we obtain a velocity of the stripped
non-thermal component of $\sim 420$\,km\,s$^{-1}$, which is consistent
with the wind velocity (V$_{ICM}$= 500\,km\,s$^{-1}$) resulting from
the $N$-body/hydrodynamical simulations of RPS affecting
SOS\,90630. The result indicates that the radio tail is produced by
the gas stripped from the disc.

The other piece of information is given by the radio spectral index
across the galaxy and the tail. Also in this case, opposite behaviours
have been reported in the literature for different
objects. \citet{Vollmer+09} did not observe a steepening of the
spectral index in the extraplanar radio emission of NGC\,4438, while
other studies \citep[][]{Vollmer+04,Chyzy+06,Chen+20} reported a rapid
ageing of the electrons with increasing distance from the galaxy
disc. For SOS\,90630 we actually observe a steepening in the inband
spectral index (Fig.~\ref{fig:90630_index}) which denotes an electron
ageing from the galaxy disc to the inner tail thus supporting that the
relativistic electrons are, at least in the inner tail, stripped from
the galaxy disc. Also considering the spectra of gal+tail1 and tail2
(see Section~\ref{radioimag}) we measure a clear steepening of the
spectral index in tail2 although corresponding to only a modest
ageing.

However, we notice that the line diagnostics analysis shows that the
excitation of the gas both within the main body of the galaxy and
within its extended tail is due to H{\sc ii} regions
\citep{ShaSSIII}. Therefore, even if we tend to interpret the radio
emission of the tail as due to electrons stripped from the galaxy
disc, in the outer tail we cannot exclude a contribution due to {\it
  in situ} star formation, which would account for the modest ageing
observed in this region.

\subsection{Origin of the relativistic electrons in the tail of SOS\,114372}

The spectra of gal+tail1 and tail2 (see Section~\ref{radioimag}) show
a steepening of the spectral index in tail2 which suggests that the
relativistic electrons origin from the galaxy disc. The ageing
observed in tail2 is nevertheless not dramatic. \citet{ACCESSV} showed
that the emission in some regions of the galaxy and in the whole gas
outflow presents a significant contribution from shock excitation,
together with emission powered by star formation. The distribution of
the regions with higher velocity dispersion and shock-ionised gas
across the galaxy shows that the ram pressure compresses and strips
the gas out of SOS\,114372, forming a tail of turbulent shock-ionised
gas and dust \citep{ACCESSV}. Shock due to ram pressure may similarly
affect and accelerate also the relativistic electrons
\citep[e.g.][]{VBK+04} and explain their modest ageing in the outer
tail.

Although the difference in length of the radio and H$\alpha$ tails may
depend to difference in sensitivity, the almost 3$\times$ longer radio
tail of SOS\,114372 deserves further investigations.

A more extended tail in radio than H$\alpha$ could be consistent with
similar results found in galaxies and explained in terms of different
time-scales involving star formation traced by the two emissions
\citep[e.g.][]{Leslie+17}. While radio emission traces star formation
on time-scales $>10$\,Myr, H$\alpha$ emission traces star formation on
smaller time-scales. \citet{Leslie+17} found that the older star
formation traced by the 1.4\,GHz emission is more extended than the
younger one traced by the H$\alpha$ emission on a sample of
star-forming galaxies. Since our highest radio frequencies
(1283-1367\,MHz, see Table~\ref{radiodata}) are very close to the
canonical radio tracer of star formation at 1.4\,GHz, the finding of
\citet{Leslie+17} could be in line with a more extended tail in radio
than H$\alpha$ in SOS\,114372 (and to a lesser extent in
SOS\,90630). Following their work, an older stellar population could
be more easily detected and more extended in the radio because it
takes time for the radio emission to be produced in supernovae and for
the cosmic rays to be transported far away from the galaxy disc along
the tails. The galaxy was also observed with MUSE by
\citet{Poggianti+19}. They observed a ionised gas tail consistent with
the radio tail length at 1283\,MHz, but still much shorter than the
radio tail observed at 943\,MHz.

Also for this galaxy, we tentatively estimate the velocity of the
stripped plasma. The radiative age ($\sim 90$\,Myr) in tail2 can be
associated to a tail length of 38\,kpc (see
Fig.~\ref{fig:masks}). After correction for the projection effects,
i.e. accounting for an angle $\phi \sim 54$\,deg between the wind
velocity and the line of sight provided by the simulations, we obtain
a velocity of stripped non-thermal component of $\sim
510$\,km\,s$^{-1}$ which is 2.7$\times$ lower than the wind velocity
(V$_{ICM}$=1400\,km\,s$^{-1}$) inferred from the
$N$-body/hydrodynamical simulations of RPS run for SOS\,114372.
However, we notice that the simulations ran by \citet{ACCESSV} did not
include galactic magnetic fields. If the ram pressure differently
affects the gas and the non-thermal component in the galaxy, it is
also possible that the radio continuum and the ionised gas tails
present different morphology which cannot be foreseen by the
simulations now available for SOS\,114372.

\subsection{The environment of the RPS galaxies}

\subsubsection{SOS\,90630 in the poor cluster SC\,1327-312}

The cluster SC\,1327-312 has a well defined centre with the peak of
the X-ray emission coinciding with a bright galaxy. The velocity
dispersion of the cluster is 535$\pm$17\,km\,$s^{-1}$
\citep{ShaSSV}. The cluster lies almost entirely within the virial
radius of A\,3558 and it remains hard to disentangle the two
clusters. In particular, SOS\,90630 is located in the outskirt of
SC\,1327-312, 0.2$r_{200}$ SW with respect to the centre.

\citet{Bardelli+02} analysed the X-ray gas profile of SC\,1327-312
showing that it is rather symmetric. However, the whole SSC shows a
continuous filamentary structure 2\,degrees in extent, that is filled
with hot gas as seen by both Planck and {\it XMM-Newton} satellite
\citep{Planck14,ACCESSV}. \cite{SOSII} identified two clear
overdensities of blue supercluster galaxies in the western region of
the poor cluster SC\,1329-313, and a linear structure bisecting the
clusters A\,3558 and SC\,1327-312. This is actually the region where
the clusters A3562 and A3558 are both experiencing merging events,
which may suggest that this merging has recently triggered star
formation in these blue galaxies, possibly through shock fronts
produced when the ICMs of merging clusters collide. Alternatively,
these overdensities could reflect infall regions along the filament
connecting the clusters A3558 and A3562. All this places SOS\,90639 in
an environment where cluster interactions occurred and/or are ongoing.

The role of the magnetic field in RPS events has been investigated by
\citet{TS14} who ran magnetohydrodynamical simulations of RPS
including a galactic magnetic field. Their work showed that the total
amount of gas stripped in the magnetised galaxies does not
significantly differ from that removed in unmagnetised ones.  They do
not find any dramatic difference in the morphology of the tails with
and without magnetic field, although the magnetic fields in the disc
lead to larger unmixed structures in the tail. Actually, we observe a
fragmented ionised gas tail in SOS\,90630.

\citet{Rus+14} simulated the interaction of disc galaxies orbiting in
a cluster and exposed to a uniformly magnetised wind including
radiative cooling and self-gravity of the gas. Such an interaction
produces a very filamentary structure in the tails which then appear
bifurcated in the plane of the sky. These features could be
distinguished in the tail of SOS\,90630 suggesting that the galaxy is
exposed to a uniformly magnetised wind.

We should mention that if the RPS explains the truncation of the gas
disc as well as other features (gas distribution and star formation in
the galaxy disc), the length of the ionised gas tail needs the help of
another force to be reproduced. \citet{ShaSSIII} suggested that both
RPS and the tidal interaction with a massive companion whose external
edge is visible in the lower right corner of
Fig.~\ref{fig:90630_r-band_radio_Ha}) contribute to the formation of
the long tail, as observed in other galaxies
\citep[e.g.][]{GBM01,VHv05}. We notice that the simulations of
\citet{ShaSSIII} assumed that a constant ram pressure is exerted on
the galaxy disc. On the other hand, \citet{T19} showed that varying
the ram pressure strength along the galaxy orbit leads to
significantly different density and velocity structure in the
tail. This new evidence given by the simulations implies that we
cannot exclude {\it a priori} that SOS\,90630 is actually affected by
RPS alone.

\subsubsection{SOS\,114372 in the rich cluster A\,3558}

The bright and massive ($M_\star = 7.7\times 10^{11}$\,M$_\odot$)
galaxy SOS\,114372 is a member of the rich cluster A\,3558 having a
velocity dispersion of 1007$\pm$25\,km\,$s^{-1}$ \citep{ShaSSV}. The
galaxy is located $\sim1$\,Mpc from the cluster centre which
corresponds to 0.4$r_{200}$, in an environment characterised by a
lower density with respect to the very cluster core. SOS\,114372
demonstrates that RPS can affect bright ($L>L^\star$) and massive
cluster galaxies both within and beyond the cluster core
\citep[e.g.][]{CK06,CK08,Chung+07} and thus showing that the ICM is
neither static nor homogeneous.

Actually, a non-homogeneous ICM is expected for non-virialized merging
and post-merging clusters as those in the SSC. From the analysis of
{\it XMM-Newton} and {\it Chandra} observations, \citet{Rossetti+07}
identified a cold front in the A\,3558 ICM probably caused by the
sloshing of the core induced by the perturbation of the gravitational
potential associated with a past merger. This cold front is expanding,
perturbing the density and the temperature of the ICM. From their
Fig.~3 it is clear that the cold front is moving in the direction of
SOS 114372 and that it is $\sim 5$\,arcmin ($\simeq 300$\,kpc in
projection) from the galaxy. Unfortunately, the X-ray observations do
not include the region of SOS 114372.

\subsubsection{ShaSS\,421: a massive galaxy member of AS\,0726}

ShaSS\,421 is a peculiar case of a massive galaxy affected by RPS in a
low-density environment. In fact, the parent clusters of ShaSS\,421,
AS\,0726, presents an extremely low X-ray luminosity
L$_{X,bol}=6.9\times10^{42}$\,erg\,s$^{-1}$ and gas temperature
T$_X=0.90$\,keV \citep[ROSAT detection,][]{DSE05}. On the other hand,
thanks to the newly acquired redshifts in this area, \citet{ShaSSV}
measured a velocity dispersion of $603\pm48$\,km\,s$^{-1}$ and the
corresponding dynamical mass of M$_{200}=2.4\times
10^{14}$\,M$_\odot$, which suggests a hotter and denser ICM with
respect to the ROSAT detection. Beside this, the dynamical analysis
revealed that AS\,0726 is a system of 2-3 groups with ShaSS\,421
associated to the most prominent peak in the galaxy spatial
distribution. Therefore, the complex and unrelaxed dynamics of
AS\,0726 could contribute to the ongoing RPS observed in
ShaSS\,421. The group-group merger would either trigger a shock front
in the ICM, or allow the high encounter velocity between ICM and
galaxy \citep[see][]{Owers+12,Venturi+22}.

A datailed analysis of this galaxy is ongoing by means of deep MUSE
data.

\subsubsection{SOS\,61086 in the poor cluster SC\,1329-313} 

SOS\,61086 is a member of the poor cluster SC\,1329-313. The galaxy
shows both H$\alpha$ and radio tails extending out to 30\,kpc north of
the disc. The tail orientation, as projected in the sky, is almost
tangential to the cluster outskirt suggesting that the galaxy orbit is
not radial.

\citet{Venturi+22} detected an inter-cluster radio bridge at GHz
frequencies which extends between the cluster A\,3562 and the group
SC\,1329-313. This bridge takes the form of an arc which connects the
cluster and the group from north, and remarkably follows the X-ray
emission in this region. The authors interpreted the observed low
brightness intercluster diffuse radio emission as a trace of
turbulence injected in the ICM by the flyby of SC\,1329-313 north of
A\,3562 into the supercluster core
\citep[see][]{Finoguenov+04}. Because of its position, the trajectory
of SC\,1329-313 falling into the SSC, and the tail orientation,
SOS\,61086 is the textbook case of a galaxy where the RPS is triggered
by the perturbed ICM during a minor merger event.

\section{Summary and conclusions}
\label{sumcon}

In this work we study a sample of four galaxies undergoing
ram-pressure stripping in different environments of the Shapley
supercluster. The radio continuum emission of three of them
(SOS\,90630, SOS\,114372 and ShaSS\,421) is analysed here for the
first time. Radio observations collected with various interferometers
and spanning the frequency range from 235 to 1367\,MHz are available
for the galaxies SOS\,90630, SOS\,114372 and SOS\,61086, while only
887\,MHz ASKAP observations cover ShaSS\,421.

SOS\,90630 and SOS\,114372 both show radio continuum extraplanar
emission in different radio bands. At 1.283~GHz MeerKAT radio
continuum tails extend $\sim 40$\,kpc in projection out of the disc,
while we do not detect a radio continuum tail in the ASKAP data of
ShaSS\,421.

Together with the total flux, in the tailed galaxies we are able to
disentangle the radio emission of the galaxy and the inner tail
(gal+tail1) from that of the outer tail (tail2). By means of the
multi-frequency dataset, we derive the spectral index along the tails,
which in both galaxies steepens in the outer tail as it was also
observed in the fourth galaxy \citep[SOS\,61086,][]{Venturi+22}.
Furthermore, we derive the inband spectral index across the disc of
SOS\,114372 and the disc and inner tail of SOS\,90630 from the
1.28\,GHz MeerKAT data. In both galaxies the tail of relativistic
electrons is superimposed to the tail of ionised gas as traced by
H$\alpha$ emission. However, at 1.28\,GHz the radio tail extends in
projection with almost the same length of the optical tail for
SOS\,90630 (as for SOS\,61086), while for SOS\,114372 the length of
radio tail significantly exceeds that of the optical one.

We interpret our findings in the framework of the RPS scenario, which
was ascertained for these galaxies by the analysis of IFS
observations, multi-band data and $N$-body/hydrodynamical simulations
\citep[][]{ACCESSV,ShaSSIII}.

In SOS\,90630 and SOS\,114372 the consistency among {\it i}) the epoch
of the onset of stripping (inferred by the hydrodynamical
simulations); {\it ii}) the age of the youngest stellar population
(derived by the IFS data) and {\it iii}) the radiative age of the
electrons in the tail (obtained from the synchrotron spectrum)
supports a scenario where the radio emitting plasma and the warm gas
tails are related to the same event, i.e. by the ram pressure, with
similar efficiency. This was also the case of SOS\,61086
\citep[][]{Venturi+22}.

{\it SOS\,90630}. Taking into account the steepening of the spectral
index, the ageing of the electrons and the source (H{\sc ii} regions)
of the gas excitation in the tail, we suggest that the radio emission
of the inner tail is due to electrons stripped from the galaxy disc,
while in the outer tail we cannot exclude a contribution due to {\it
  in situ} star formation, which would account for the modest ageing
observed in this region.

{\it SOS\,114372}. The steepening of the spectral index indicates that
also in this galaxy the relativistic electrons origin from the disc,
however the H$\alpha$ emission in the tail is mainly powered by shock
excitation which may accelerate also the relativistic electrons and
explain their modest ageing in the outer tail.

Considering also SOS\,61086, the three detected radio tails are all
fuelled by relativistic electrons stripped with the plasma from the
disc. However, while the electrons in the tail of SOS\,61086 do not
show any signs of re-acceleration or rejuvenation, those in the tails
of the other two galaxies are likely either accelerated by shock
excitation (SOS\,114372) or further powered by {\it in situ} star
formation (SOS\,90630).

It is remarkable that all the galaxies of our sample are located in
environments where cluster-cluster interaction occurred and/or is
ongoing. We underline again that the galaxies considered in this work
are the most dramatically affected by RPS across the ShaSS region.

The correlation between clusters and groups merger and RPS events has
been suggested by \citet{OCN12} who identified three RPS galaxies
showing active star formation in their tails in Abell\,2744. Being
these objects found in close proximity to the merger-related feature,
the authors proposed that the observed star formation has been
triggered by the rapid increase in pressure during an interaction with
the shock front. \citet{McPartland+16} conducted a systematic search
for RPS galaxies in 63 MACS clusters at intermediate redshifts
identifying 53 RPS candidates. Their analysis of a three-dimensional
modelling of the orbits of the galaxies suggests that the most extreme
events are primarily triggered in massive cluster mergers where the
RPS may initiate far from the cluster core because of the large
relative (galaxies-ICM) velocities. This could be actually the case of
SOS\,114372. In addition, \citet{McPartland+16} suggested that extreme
RPS can also occur in mergers of poor clusters and groups. This is
likey what we observe for ShaSS\,421 in the unrelaxed poor cluster
AS\,0726.

A first theoretical study of RPS in the environment of galaxy cluster
merges was presented by \citet{Ruggiero+19} considering the A901/2
system where two clusters and two groups are simultaneously
merging. They concluded that the galaxies experiencing RPS are
preferentially located near a boundary where the gas moving with the
cluster/subcluster and the gas from the remaining of the system
collide. This recalls the case of SOS\,61086 affected by RPS along the
trajectory of the poor cluster SC\,1329-313 falling into the SSC.

If cluster-cluster interaction plays a role in favouring the RPS, as
it seems, then the detection of RPS events can be used as a diagnostic
tool to constrain the geometry of cluster collision as proposed by
\citet{EK19}. Actually, the orientation of the tail of SOS\,61086
supports and reinforce the scenario of the minor merger of
SC\,1329-313 and A\,3562 in the SSC, as recently traced by the
intercluster diffuse emission detected at 1.28\,GHz by MeerKAT
\citep[see][]{Venturi+22}.

The question now is why our sample is limited to four galaxies
experiencing RPS across the harsh environment of the Shapley
supercluster. These ascertained cases of ongoing RPS have been
selected from the optical images, thus biased to objects showing star
formation and/or shock excitation of the gas into the tail. Different
biases, both observational and intrinsic to the mechanism (e.g. the
short duration of the stripping event), make it difficult to identify
RPS candidates and deep/telescope time expensive IFS observations are
needed to trace low brightness extraplanar ionised gas. Therefore,
even if the information provided by IFS observations will be always
essential, the current available and forthcoming radio facilities
(MeerKAT, LOFAR, uGMRT, ASKAP) will allow to map the atomic gas
content and detect extra-planar atomic hydrogen as well as to detect
tails of non thermal component by means of blind surveys of large sky
regions \citep[e.g.][]{Roberts+21I}. We aim to use this approach to
make a census of RPS galaxies and possibly trace their radio continuum
and H{\sc i} tails in the Shapley supercluster.

\section{Acknowledgements}

The authors thank the referee for helping them to improve their work.
PM, TV, VC and GB acknowledge INAF Mini Grant 2022 {\it ShaSEE -
  Shapley Supercluster Exploration and Exploitation}. Basic research
in radio astronomy at the Naval Research Laboratory is supported by
6.1 Base funding. This work was supported in part by the Italian
Ministry of Foreign Affairs and International Cooperation, grant
number ZA23GR03. We thank the staff of the GMRT that made these
observations possible. GMRT is run by the National Centre for Radio
Astrophysics of the Tata Institute of Fundamental Research. This
scientific work uses data obtained from Inyarrimanha Ilgari Bundara /
the Murchison Radio-astronomy Observatory. We acknowledge the Wajarri
Yamaji People as the Traditional Owners and native title holders of
the Observatory site. CSIRO ASKAP radio telescope is part of the
Australia Telescope National Facility
(https://ror.org/05qajvd42). Operation of ASKAP is funded by the
Australian Government with support from the National Collaborative
Research Infrastructure Strategy. ASKAP uses the resources of the
Pawsey Supercomputing Research Centre. Establishment of ASKAP,
Inyarrimanha Ilgari Bundara, the CSIRO Murchison Radio-astronomy
Observatory and the Pawsey Supercomputing Research Centre are
initiatives of the Australian Government, with support from the
Government of Western Australia and the Science and Industry Endowment
Fund. The MeerKAT telescope is operated by the South African Radio
Astronomy Observatory, which is a facility of the National Research
Foundation, an agency of the Department of Science and
Innovation. Based on observations collected at the European
Organisation for Astronomical Research in the Southern Hemisphere
under ESO programmes 097.B-0917(A) and 099.B-0016(A).

\section*{Data Availability}

In this work we exploit the radio continuum observations of the
Australian Square Kilometre Array Pathfinder
\citep[ASKAP,][]{Hotan21}. In particular, at 887\,MHz we used the data
of the survey EMU Evolutionary Map of the Universe
\citep[][http://emu-survey.org/]{Norris11} together with the Early
Science Project ESP\,20 \citep{Venturi+22}, and at 943\,MHz and
1367\,MHz the data of the POSSUM \citep[Polarization Sky Survey of the
  Universe's
  Magnetism,][https://possum-survey.org/science/]{Gaensler+10}
Pilot\,2 survey. We also benefit of one pointing of the MeerKAT Galaxy
Cluster Legacy Survey \citep[MGCLS,][]{Knowles+22} at 1283\,MHz. The
VST $r$-band images of the galaxies have been collected and analysed
in the Shapley Supercluster Survey \citep[ShaSS,][]{ShaSSI}.  The raw
images are available through the ESO archive
(http://archive.eso.org/cms.html), ESO Programmes 096.A-0129
091.A-0050, 089.A-0095, 090.B-0414. The reduced $r$-band images can be
provided by the authors upon request. The ESO MUSE observations of
ShaSS\,421 (ESO Programme 097.B-0917) will be published by Merluzzi et
al. (in preparation).




\bibliographystyle{mnras}
\bibliography{biblio_RPSs} 




\appendix

\section{Full resolution and convolved radio images of the galaxies SOS\,90630 and
  SOS\,114372}
\label{AppA}

\begin{figure*}
\begin{centering}
\includegraphics[width=160mm]{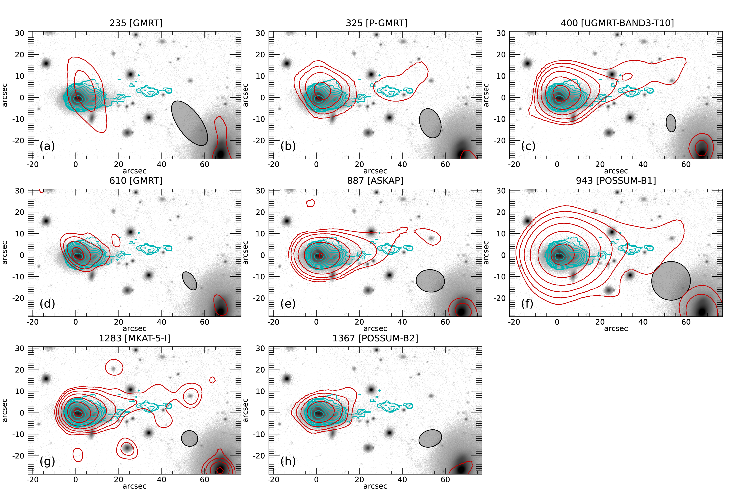}
\end{centering}
\caption{Full resolution radio images (red contours) in all the bands
  for the galaxy SOS\,90630 superimposed to the VST $r$-band image. On
  top of each panel the frequency (MHz) and instrument are indicated
  (see Table~\ref{radiodata} for details). The contours start from
  3$\times$rms and are spaced by a factor 2 in flux. The cyan contours
  map the H$\alpha$ emission. The beam is represented by the gray
  ellipse.
\label{fig:x8_090630}}
\end{figure*}

\begin{figure*}
\begin{centering}
\includegraphics[width=160mm]{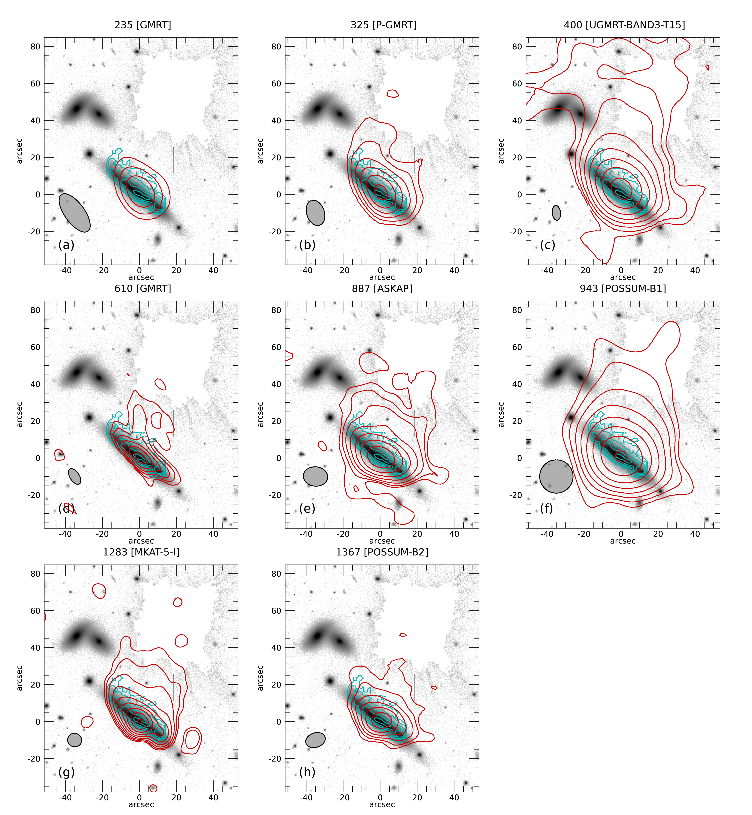}
\end{centering}
\caption{Same of Fig.~\ref{fig:x8_090630} for SOS\,114372.
\label{fig:x8_114372}}
\end{figure*}

\begin{figure*}
\begin{centering}
  \includegraphics[width=160mm]{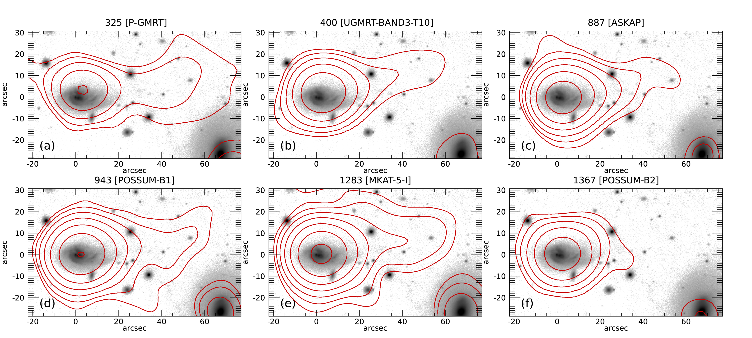}
\end{centering}  
\caption{Radio contours (red) convolved to 18$\times$18\,arcsec$^2$ in
  all the bands for the galaxy SOS\,90630 superimposed to the VST
  $r$-band image. On top of each panel the frequency (MHz) and
  instrument are indicated (see Table~\ref{radiodata} for
  details). The contours start from 3$\times$rms and are spaced by a
  factor 2 in flux.
\label{fig:x3_090630}}
\end{figure*}

\begin{figure*}
\begin{centering}
\includegraphics[width=160mm]{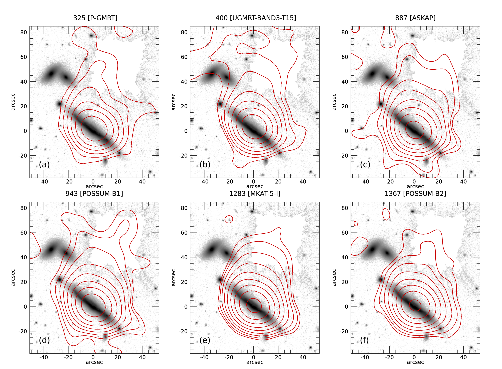}
\end{centering}
\caption{Same of Fig.~\ref{fig:x3_090630} for SOS\,114372.
\label{fig:x3_114372}}
\end{figure*}

In this section we show the radio images of the two galaxies newly
studied in this work and for which multi-frequency radio observations
are available (see Table~\ref{radiodata} for the angular resolution
and rms of each image). The full resolution images are shown in
Figs.~\ref{fig:x8_090630} and \ref{fig:x8_114372}, while the images
convolved to 18$\times$18\,arcsec$^2$ are shown in
Figs.~\ref{fig:x3_090630} and \ref{fig:x3_114372}.

\section{Spectral fits}
\label{AppB}

\begin{figure*}
  \begin{centering}
\includegraphics[width=160mm]{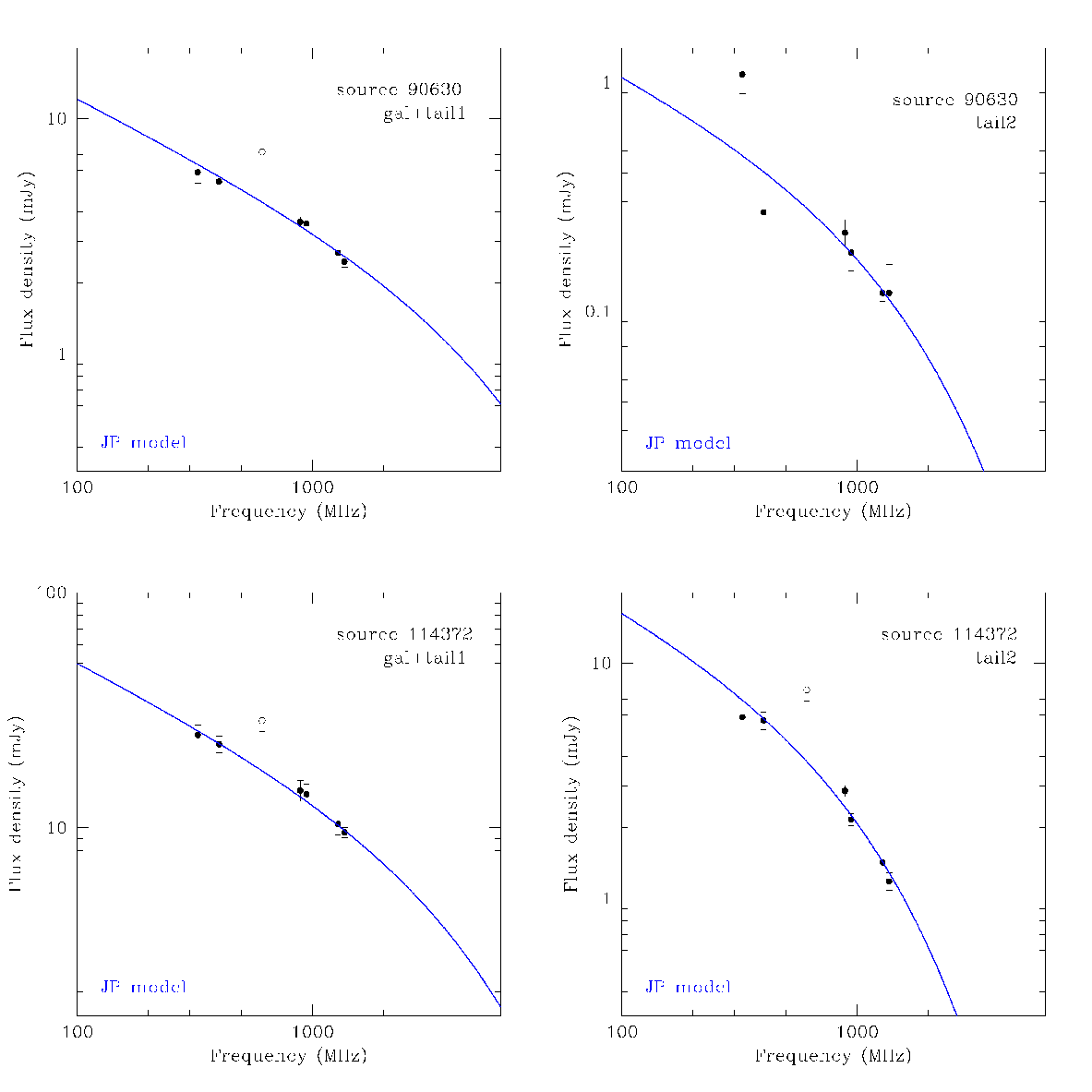}
\end{centering}
\caption{{Spectral fits obtained for the component gal+tail1 and tail2
    using the information provided in Table~\ref{radiofluxes18} for
    SOS\,90630 (top row) and for SOS\,114372 (bottom row). The fluxes
    at 610\,MHz (open circle) were not considered in the fits.}
\label{fig:spectralfits}}
\end{figure*}

In Table~\ref{radiofluxes18} we give the fluxes measured from the
radio images convolved to the resolution of 18$\times$18\,arcsec$^2$
except GMRT 235\,MHz for SOS\,90630.  In Fig.~\ref{fig:spectralfits}
we plot the spectral fits to the data in Table~\ref{radiofluxes18} for
SOS\,90630 and SOS\,114372 obtained with the program Synage
\citep{Murgia+99}. The parameters used for the fit are detailed in
Sections~\ref{RT_90630} and \ref{RT_114372}.

\section{$N$-body/hydrodynamical simulations of the RPS galaxies}
\label{AppC}

The main goal of the simulations is to understand to what extent ram
pressure might explain the main observed features of the gas: the
truncation of the disc, the shape and extension of the tail and the
kinematics. We ran N-body/hydrodynamical simulations for the galaxies
SOS\,114372, SOS\,90630 and SOS\,61086 in \citet{ACCESSV,ShaSSIII} to
which we refer for a complete description. The aim of the simulations
was to reproduce as closely as possible the observed velocity field of
the gas and thus constrain the ICM wind angle and velocity and the
time of the onset of ram pressure. The simulations were run from wind
angles close to edge-on to close to face-on at intervals of 10$^\circ$
and the wind velocity range was set taking into account as the lower
limit the galaxy line-of-sight velocity relative to the cluster
systemic velocity and as the upper limit the escape velocity (derived
from the Hernquist model of each cluster). The ICM densities were
derived from the X-ray profile of each cluster \citep{SP10,ShaSSIII}.

For the three galaxies the simulations were performed adopting a model
galaxy resembling the pre-interaction properties of each galaxy. The
initial conditions are calculated according to \citet{Springel+05}
based on the work of \citet{Mo+98}. Our model galaxy includes a dark
matter halo, a stellar disc, a gaseous disc and a stellar bulge. The
disc scale length of the galaxy is directly related to the angular
momentum of the disc with the spin as a free parameter. The stellar
and gaseous disc have exponential surface density profiles. For the
stellar bulge and the dark matter halo a \citet{Hernquist90} profile
is adopted.

The simulations were done with the cosmological $N$-body/hydrodynamic
code GADGET-2 \citep{Springel05} for SOS\,114372 while the
cosmological moving-mesh code AREPO \citep{Springel10} is used for
SOS\,90630 and SOS\,61086 \citep[for details see][]{ACCESSV,ShaSSIII}.



\bsp	
\label{lastpage}
\end{document}